\documentstyle[12pt]{article}

\font\bigg=cmbx10 at 22truept     at 12 truept
\def\be{\begin{equation}}
\def\ee{\end{equation}}
\def\s{\scriptstyle}
\def\ss{\scriptscriptstyle}

\def\ds{\displaystyle}

\def\o{\over}
\def\C{{\s C}}

\def\pmic{P(\C_i,t)}

\def\pmicl{P(\C_i,t+1)}

\def\pmicf{P'(\C_i,t)}
\def\pmicjf{P'(\C_j,t)}
\def\pmicc{P_c(\C_i,t)}
\def\pmicjc{P_c(\C_j,t)}

\def\pmicLf{P'(\C_i^{\ss L},t)}
\def\pmicRf{P'(\C_i^{\ss R},t)}

\def\fb{\bar f}        \def\fav{{\bar f}(t)}

\def\fmic{f(\C_i,t)}

\def\muti{{\cal P}({\ss C_i})}

\def\mutji{{\cal P}({\ss C_j\ra C_i})}

\def\crossij{{\cal C}_{\ss C_iC_j}^{(1)}(k)}
\def\crossjl{{\cal C}_{\ss C_jC_l}^{(2)}(k)}
\def\hamij{d^H(i,j)}
\def\hamijL{d^H_L(i,j)}
\def\hamijR{d^H_R(i,j)}
\def\hamilL{d^H_L(i,l)}
\def\hamilR{d^H_R(i,l)}
\def\ra{\rightarrow}  \def\lra{\longrightarrow}

\def\ef{f_{\ss\rm eff}({\ss C_i},t)}

\def\nn{\nonumber}

\begin{document}

  \vspace{1.5cm}
  \begin{titlepage}
\begin{flushright}
   {ICN-UNAM-97-10}\\
   {1st July, 1997}\\ 
\end{flushright}
\vskip 0.3truein
  \begin{center}
  {\bigg Self-Adaptation in Evolving Systems}\\
  \vspace{1cm}
{\large C.R. Stephens$^1$,  I. Garc\'\i a Olmedo$^2$, J. Mora Vargas$^3$ }\\
{\large and H. Waelbroeck$^1$}\\
\vspace{1.5cm}
  $^1$ {\em  Instituto de Ciencias Nucleares, UNAM\\
  Circuito Exterior, A. Postal 70-543,\\
   M\'exico D.F. 04510}\\
\vskip 0.1in
  $^2${\em  Lab. de Visualizaci\'on, DGSCA, UNAM\\
  Circuito Exterior, C.U.\\
   M\'exico D.F. 04510}\\
\vskip 0.1in

  $^3${\em  Facultad de Ingenier\'\i a, UNAM\\
  Circuito Exterior, C.U.\\
   M\'exico D.F. 04510}\\
  \end{center}
  \vspace{2cm}
\begin{abstract}  
A theoretical and experimental analysis is made of the effects of self-adaptation 
in a simple evolving system. Specifically, we consider the effects of coding the 
mutation and crossover probabilities of a genetic algorithm evolving in certain
model fitness landscapes. The resultant genotype-phenotype mapping is degenerate,
there being no direct selective advantage for one probability versus another. We
show that the action of mutation and crossover breaks this degeneracy leading to 
an induced symmetry breaking among the genotypic synonyms. We demonstrate that
this induced symmetry breaking allows the system to self-adapt in a time 
dependent environment.

\end{abstract}
\end{titlepage}

\section{Introduction}
One of the most striking features of complex systems, especially in the 
biological realm, is the ability to adapt. Loosely speaking, this means
``optimization'' in a time and/or position dependent ``environment''.
The Darwinian paradigm of natural selection offers an intuitive 
framework within which one may try to understand such adaptive 
behaviour. Recently, another alternative, or rather complementary, 
paradigm has been suggested \cite{kauffman} that takes as its 
principal theme --- ``spontaneous ordering''. In this case 
advantageous or disadvantageous characteristics may appear 
via a ``spontaneous'' symmetry breaking. The principal theme of
this paper is to show that adaptation can also occur due to an
induced symmetry breaking --- induced by the action of the 
genetic operators other than selection. 
We will demonstrate the above in the context of the evolution
of a genetic algorithm (GA) seen as a simple, artificial
model of an evolving, adaptive complex system. 

Evolutionary algorithms, and in particular GAs, have played
an increasingly important role in a wide variety
of problems (see for example \cite{1} for a recent review). Two crucial
ingredients of such algorithms are: i) the existence of a population
of bit strings/chromosomes, each one of which codifies directly or
indirectly a possible solution to a problem; and ii) a set of genetic
operators which act on the strings. 
In the case of GAs the most popular and most studied 
operators are selection, mutation and crossover, though of course in 
principle the list is infinite. Many other evolutionary algorithms, such as 
evolutionary programming \cite{fogel}, rely solely on selection and mutation. 
It has been asserted that GAs are better at function optimization than 
evolutionary algorithms that rely solely on mutation 
and that this advantage is due to crossover \cite{davis}.

A lot of effort has been spent trying to understand the differing roles of
selection, mutation and crossover in the context of different fitness 
landscapes, and in particular on discovering what constitute optimal
values for the exogenous parameters, e.g. mutation probability, 
crossover probability etc. \cite{param}. The chief motivation behind this work 
is to find parameter values which maximise performance over a wide class
of test functions. However, it has been shown \cite{hart} that effectively
the most efficient parameter settings must depend on the specific 
function. 

Intuitively it is clear that time dependent parameters should
be utilized as will be illustrated clearly in this paper. In this 
regard it is possible to
specify a priori a schedule for the parameter changes, as is done in 
simulated annealing with a cooling schedule. A more attractive 
possibility, however, is to code the parameters within the 
chromosomes themselves and allow the system to ``self-optimize'' via
a genetic search through various possible parameter values. We will
refer to the continual adjustment of these parameters according to 
changes in the landscape as ``self-adaptation''.
Such self-adaptation first appeared in the context of evolution 
strategies \cite{schwefel} and has been studied by B\"ack and
collaborators \cite{back} in both the context of evolution strategies 
and GAs. One of the prime points of interest was the derivation of
optimal mutation rates on a string by string basis in the case of 
certain simple landscapes. Crossover has been considered in 
\cite{mori} where the actual crossing points were codified in the 
strings. Codification of the crossover probability itself was not 
considered.

In this paper we will consider the effects of parameter codification 
of a GA in a set of model fitness landscapes. Our principal aim is
not to show that parameter codification can lead to much better
function optimization capabilities in GAs, though this may well be the 
case, but rather to study how self-adaptation occurs in a simple
complex system. This self-adaptation will be shown to come about
not via the mechanism of Darwinian natural selection but via an 
induced symmetry breaking.

\section{Optimal Mutation and Crossover Rates}

In this section we will make some observations about optimal parameter
settings based on an analysis of the evolution
equation, valid in the large population limit, for GAs derived in 
\cite{stewael} that takes the following form
\be
\pmicl=\muti\pmicc \nn\\
+ {\ds\sum_{\ss C_j\neq C_i}}\mutji\pmicjc
\label{mastcrossmut}
\ee
where
\begin{eqnarray}
\pmicc & = & \pmicf
-{p_c\o N-1}{\ds\sum_{\ss C_j\neq C_i}}{\ds\sum_{k=1}^{\ss N-1}}
\crossij\pmicf\pmicjf \nn \\ 
& &{} +{p_c\o N-1}\sum_{\ss C_j\neq C_i}\sum_{\ss C_l\neq C_i}\sum_{k=1}^{\ss N-1}
\crossjl\pmicjf P'(\C_l, t)
\end{eqnarray}
$p_c$ is the probability to implement crossover in the first place, $k$ is the crossover point, 
and the coefficients $\crossij$ and $\crossjl$ are given by
\be
\crossij=\theta(\hamijL)\theta(\hamijR)\label{thetas}
\ee
and
\be
\crossjl={1\o2}[\delta(\hamijL)\delta(\hamilR)\nn\\
+\delta(\hamijR)\delta(\hamilL)]\label{deltas}
\ee
where $\hamijR$ is the Hamming distance between the right halves of the strings $C_i$ and
$C_j$,  ``right'' being defined relative to the crossover point. 
The other quantities in (\ref{thetas}) and (\ref{deltas}) are defined analogously. 
$\theta(x)=1$ for $x>0$ and
is $0$ for $x<0$, whilst $\delta(x)=0\ \ \forall x\neq0$ and $\delta(0)=1$. 
$\pmicf=(\fmic/\fav) P(C_i,t)$ with
$\fmic$ being the fitness of string $\C_i$ at time $t$,  and
$\fav=\sum_i\fmic\pmic$ is the average string fitness. 
The effective mutation coefficients are
$\muti=\prod_{\s k=1}^{\ss N}(1-p(k))$, which is the probability that 
string $i$ remains unmutated, $p(k)$ being the probability of mutation 
of bit $k$, which we assume to be a constant, though the equations are 
essentially unchanged if we also include a dependence on time. 
$\mutji$ is the probability that string $j$ is mutated into string $i$,
\be
\mutji=\prod_{\ss{k\in \{C_j-C_i\}}}{p(k)}\prod_{\ss{k\in\{C_j-C_i\}_c}}
{(1-p(k))}
\ee
where $\s\{C_j-C_i\}$ is the set of bits that differ between $\C_j$ and 
$\C_i$ and $\s\{C_j-C_i\}_c$, the complement of this set, is the set 
of bits that are the same. In the limit where the mutation
rate $p$ is uniform, $\muti=(1-p)^{\s N}$ and 
$\mutji=p^{\s d^{\ss H}(i,j)}(1-p)^{\s N-d^{\ss H}(i,j)}$,
where $\hamij$ is the Hamming distance between the strings $\C_i$ and $\C_j$.
$\crossij$ is the probability that given that $\C_i$ 
was one of the parents it is destroyed by the crossover process. 
$\crossjl$ is the probability that given that neither parent was $\C_i$ it 
is created by the crossover process. 

This evolution equation takes into account 
exactly, given the approximation of a large population, the effects of destruction 
and reconstruction of strings. By realizing that string reconstruction via crossover 
takes place via the concatenation of two strings, one of which reconstructs the left
part of the desired string, $C_i^L$, and the other the right part, $C_i^R$, 
one can write the equation in a much more elegant form by using
\be
\pmicc=\pmicf \nn\\
-{p_c\o N-1}\sum_{k=1}^{\s N-1}(\pmicf-\pmicLf\pmicRf)\label{stringfin}
\ee
with 
\be\pmicLf=\sum_{\ss C_j\supset C_i^L}\pmicjf\ee
and similarly for $\pmicRf$.  

We may now use the above equation to investigate the evolution of any function. 
One of particular interest is the increment in average population fitness per generation
\be
\delta\fb=\fb(t+1)-\fb(t)\label{2}
\ee
which using the evolution equation (\ref{mastcrossmut}) we can write as
\be
\delta\fb={\ds\sum_{\ss C_i}}\fmic\muti\pmicc \nn\\
+ {\ds\sum_{\ss C_i}}{\ds\sum_{\ss C_j\neq C_i}}\fmic\mutji\pmicjc
-{\ds\sum_{\ss C_i}}\fmic\pmic\label{delta}
\ee
Note that this equation is highly non-linear in the mutation rate $p$, but
linear in the crossover probability $p_c$. 

The values of $p$ and $p_c$
play a very important role in determining the success of a GA. The
specific values required in a particular problem however will depend on what
we mean by success. There are several different ways in which one may 
evaluate the success of a particular algorithm. One would be to maximise the
average population fitness, $\fb$, generation by generation. Another would be
the success probability \cite{back}
\be
S^+(\{p_i\})={\cal P}(f(p(g))>f(g))
\ee
where $\{p_i\}$ represents the genetic operator probabilities and $p(g)$ is 
the image of the genotype $g$ under the action of all genetic operators.
Of course, one can restrict the definition of the success probability to
be associated with a particular genetic operator. 

Here we will concentrate on the average fitness improvement. As $\delta\fb$
depends only linearly on $p_c$, the value of $p_c$ that optimizes fitness growth
over one time step, if we neglect finite size effects, will be either $0$ or $1$ . This is 
somewhat deceptive however in that the integrated fitness gain, $\fb(t)-\fb(0)$, will be
highly non-linear in $p_c$ except when $t=1$. For a given time step however, from
(\ref{stringfin}) one can see that whether or not $p_c$ should be minimized or
maximized depends on whether parts of highly fit strings are positively or 
negatively correlated in the selected parent population. If the correlation is negative,
i.e. $\pmicf<\pmicLf\pmicRf$,
then the reconstruction of highly fit strings dominates over destruction and $p_c$
should be maximized. To illustrate this we restrict attention for the moment 
to the case $p=0$. One finds
\be
\delta\fb=\Delta\fb
-{p_c\o N-1}\sum_{k=1}^{\s N-1}{\ds\sum_{\ss C_i}}\fmic
(\pmicf-\pmicLf\pmicRf)\label{fitcross}
\ee
where $\Delta\fb(t)=({\bar {f^2}}-{(\fb)}^2)/\fb$ is the variance in the population 
fitness. As is well known $\Delta\fb(t)\geq 0$, i.e. fitness in the absence of 
other genetic operators is a Lyapunov
function. As stated, whether or not crossover helps or hinders fitness growth
depends on whether reconstruction or destruction of highly fit strings is dominant.

We can get some idea of when this occurs by restricting attention to the 
very simple case of a population of two bit strings (one can also think of this
in terms of two schematas). One finds
\be
\delta\fb=\Delta\fb
-{p_c\o \fb^2}(f_{\ss 00}+f_{\ss 11}-f_{\ss 10}-f_{\ss 01})
(f_{\ss 10}P_{\ss 10}f_{\ss 01}P_{\ss 01}-f_{\ss 00}P_{\ss 00}f_{\ss 11}P_{\ss 11})
\label{deltaff}
\ee
For crossover to be a net positive force we require either
\be 
f_{\ss 00}+f_{\ss 11}>f_{\ss 10}+f_{\ss 01} \ \ \ \ \ {\rm and} \ \ \ \ \ 
f_{\ss 10}P_{\ss 10}f_{\ss 01}P_{\ss 01}>f_{\ss 00}P_{\ss 00}f_{\ss 11}P_{\ss 11}
\ee
or
\be
f_{\ss 00}+f_{\ss 11}<f_{\ss 10}+f_{\ss 01} \ \ \ \ \ {\rm and} \ \ \ \ \ 
f_{\ss 10}P_{\ss 10}f_{\ss 01}P_{\ss 01}<f_{\ss 00}P_{\ss 00}f_{\ss 11}P_{\ss 11}
\ee
For a linear fitness landscape, as $f_{\ss 00}+f_{\ss 11}=f_{\ss 10}+f_{\ss 01}$,
crossover in this $2$-bit problem is neutral, although we once again emphasize that this is
at the level of one time step. An interesting corollary of this is that even in the presence
of crossover $\fb$ is a Lyapunov function. 
However, for a deceptive landscape
of type I or type II \cite{goldberg} one finds that the effect of crossover is destructive.
In this case having low values of $p_c$ would be beneficial. Of course if finite size 
effects are important having $p_c$ too low may inhibit genetic diversity.

We can try to generalize this lesson beyond the case of two-bit strings. In (\ref{fitcross})
one may see that in the sum over strings, if the landscape is deceptive 
in the sense that $\pmicLf\pmicRf<\pmicf$, then crossover will inhibit fitness growth;
whilst if $\pmicLf\pmicRf>\pmicf$ it will enhance fitness growth. In the former case
a low $p_c$ would be beneficial whilst in the latter a high value. Generally speaking,
the more deceptive the landscape the lower the optimum value of $p_c$. In this sense
GA hard problems may be made GA easier by a reduction in the crossover probability.
Note also that we may have deception with respect to one crossover point but not
with respect to another; this would imply that a crossover probability, $p_c(k)$, 
that depends on the crossover point might well be useful.
The point of the above is to show that what constitutes an optimal value for crossover is
very much landscape dependent, moreover it may very well be time dependent in that
at one time the population may be much more deceptive than at another time.

We now turn our attention to mutations. Once again for simplicity we will turn off the
other genetic operator and consider mutations in isolation. We will also assume that
$p$ is constant. In this case 
\be
\delta\fb={\ds\sum_{\ss C_i}}{\ds\sum_{\ss C_j}}\fmic\pmicjf p^{\hamij}(1-p)^{N-\hamij}
-{\ds\sum_{\ss C_i}}\fmic\pmic
\ee
thus
\be 
{d\delta\fb\o dp}=
{\ds\sum_{\ss C_j}}{\ds\sum_{\ss C_j}}\fmic\pmicjf p^{\hamij-1}(1-p)^{N-\hamij-1}(\hamij-pN)
\label{mutopt}
\ee
We now wish to find the extrema. Doing the explicit sums over ${\s C_i}$ is of course
very difficult. One solution is to choose $p^*=\hamij/N$, however, this requires
making $p$ string specific. One can get an estimate of an optimum $p$ by solving
$d\delta\fb/ dp=0$ in a mean field approximation where we replace $\hamij$ by
$<\hamij>$, its expectation value over the population. This results in $p^*=<\hamij>/N$.
For a random population, where $<\hamij>\sim N/2$, $p^*\sim 1/2$ which accords with
intuition. Near the ordered limit where $<\hamij>\sim 0$, $p^*\sim 0$ which once again
accords with intuition. Clearly however $p^*$ will be time dependent. Note that 
whether or not zero mutation rate in the ordered limit is optimal or not will depend
on whether the population has ordered about the global optimum or a local optimum. 
In the case of the latter $p=0$ will hinder rather than help. 

The point of the above is that different values of the GA parameters can lead to very
different performances. Trying to determine optimal parameter values empirically is
extremely difficult therefore one is naturally led to the idea of allowing the GA to
optimize itself via a codification of the parameters. 

\section{Parameter Codification and Symmetry Breaking}

Parameter codification can be done in various ways. Here we choose the simple 
route of extending the size of the chromosome such that a part of the chromosome
now carries information on the mutation and/or crossover probabilities. We assume 
that the fitness of a chromosome is not affected directly by the values of the 
parameters. This means that the genotype--phenotype mapping is now non-injective.

Consider the different classes of maps that may be defined: 
first, $f_G:G\lra R^+$ where $G$ denotes 
the space of genotypes and $f_{\ss G}$ is the fitness function that assigns a number
to a given genotype; second, $f_{\ss Q}:\lra R^+$, where 
$Q$  is the space of phenotypes. It should be emphasized that these mappings 
may also be explicitly time dependent. In fact this will normally be the case 
when the ``environment'' is time dependent. The maps may be injective or surjective. If
they are non-injective then there exist ``synonomous'' genotypes or phenotypes, i.e.
there is ``redundancy'' in the mapping. If we assume there exists a map 
$\phi:G\lra Q$ between genotype and phenotype then we have $f_{\ss G}=f_{\ss Q}\circ\phi$,
i.e. the composite map induces a fitness function on $G$. 
The map $\phi$ we may fruitfully think of as being an ``interpreter'', in that 
the map translates the genotypic information into something we call the phenotype,
where generically the fitness function will have a more intuitive interpretation.
  
One of the principle reasons for using an interpreter in GAs, 
is that, given a set of genetic operators, the original genotypic coding may not be the 
most efficient. This is the case for a binary coding in the standard scenario where
selection, mutation and simple crossover are the preferred operators. It has been 
found that a Gray coding \cite{Wright} that maps Euclidean neighbourhoods into
Hamming neighbourhoods is more effective \cite{Caruna}.  
 
In the problem at hand, where we are considering codification of parameters, 
$G$ is simply the set of chromosomes, including the parts that code for the
parameter values. $Q$ is now the space of truncated chromosomes where the genes
that code for the different parameter values have been removed. The interpreter here
is clearly non-injective. If we use an $n_c$-bit binary codification of the parameters
then the interpreter, and hence the genotypic fitness landscape, 
will have a $2^{\s n_c}$-fold degeneracy, i.e. for every phenotype there 
will be $2^{\s n_c}$ corresponding genotypes. 

This genotypic symmetry would be broken spontaneously in a finite breeding pool, 
by the theory of branching processes,
this observation being the backbone of the Neutral Theory of molecular 
evolution \cite{Kimura}. However, there 
will also be an {\it induced symmetry breaking} from the violation 
of the synonym symmetry by the genetic operators themselves. 

 If one considers the growth of a particular bit over time 
selection forces will take into account not only
the selective advantage of this bit but also its ability to produce 
well-adapted offspring, which can themselves produce well-adapted offspring,
etc. Since mutation and crossover act differently on synonymous bits, 
the synonyms will differ in their descendence, both in the passive 
sense of surviving mutations, and in the active 
sense of generating new genetic solutions. This implies that the time-averaged 
{\it effective fitness} function, defined as the growth rate of a bit over 
many generations, does not respect the synonym symmetry.
Thus, the effective fitness function provides a selective pressure which enhances the 
production of potentially successful mutants by selecting, among the synonyms, 
those that have a higher probability to generate well-adapted offspring.

We will now discuss in the context of a very simple model the phenomenon of induced 
symmetry breaking. 
 The model we consider consists of three possible genotypes, $A $, $B$ and $ C$, where
each genotype can mutate to the two adjacent genotypes. $A$ and $C$ are synonyms 
in that they encode the same phenotype $a$, i.e. $\phi(A)=\phi(C)=a$, whilst $B$ encodes
the phenotype $b$. 
In a random population, $p(A) = p(B) = p(C) = {1 \over 3}$. 
If there is probability $\mu_1/2$ for $A$ to mutate to $B$ or $C$ and probability $\mu_2/2$ for
all other possible mutations then the evolution equation that describes this 
system in the large population limit is
\be
P_i(t+1)=(1-\mu_i)P'_i(t)+(\mu_{i-1}P'_{i-1}(t)+\mu_{i+1}P'_{i+1}(t))
\ee
$P_i(t)$ being the population fraction of genotype $i$. $\mu_i$ is the probability of 
mutation from genotype $i$ to any other genotype, whilst $\mu_{i-1}$ and $\mu_{i+1}$ 
are the mutation rates from genotypes $(i-1)$ and $(i+1)$ to genotype $i$ respectively.
If we assume a simple fitness landscape, $f_a=2$, $f_b=1$, then for $\mu_i=0$ the steady state
population is $P(A)=P(C)=1/2$ and $P(B)=0$. Thus we see the synonym symmetry
is unbroken. However, for $\mu_i>0$, the genotype distribution at $t=1$ starting from a 
random distribution at $t=0$ is: $P(A)=(4-4\mu_1+3\mu_2)/10$, $P(B)=(1+\mu_1)/5$, 
$P(C)=(4+2\mu_1-3\mu_2)/10$. Thus we see that there is an induced breaking 
of the synonym symmetry due to the effects of mutation. The full effect can be dramatically
seen in Figure 1 where we have $\mu_1=0.1$, $\mu_2=0.01$. Clearly, the less mutable
string $C$ is strongly favoured over the other synonym $A$.

The effective fitness \cite{stewael} defined via
\be
P_i(t+1)={f_{\ss eff_i}(t)\over {\bar f}(t)}P_i(t)
\ee
is from the evolution equation (\ref{mastcrossmut}) given by
\be
\ef={\fb\o \pmic}\left(\muti\pmicc 
+ {\ds\sum_{\ss C_j\neq C_i}}\mutji\pmicjc\right)
\ee
Explicitly in this model
\be
f_{\ss eff_i}(t) = f_i+{1\over P_i(t)}(\mu_{i-1}f_{i-1}P_{i-1}(t)+
\mu_{i+1}f_{i+1}P_{i+1}(t)-\mu_if_iP_i(t))
\ee
At $t=0$, $f_{\ss eff_A}(0)=(4-4\mu_1+3\mu_2)/2$, 
$f_{\ss eff_B}(0)=(2+2\mu_1-\mu_2)/2$ and
$f_{\ss eff_C}(0)=(4+4\mu_1-3\mu_2)/2$. For the case 
$\mu_1=0.1$, $\mu_2=0.01$;  $f_{\ss eff_A}(0)=1.815$, $f_{\ss eff_B}(0)=1.095$
and $f_{\ss eff_C}(0)=2.195$. Thus as mentioned above we see that the 
effective fitness function provides a selective pressure by selecting among the 
synonyms those that have a higher probability to produce fit descendents.  

\section{Methodology}

In this section we will discuss the model problems we chose to illustrate 
the phenomenon of parameter adaptation. We will show that parameter 
codification yields significantly different results to those of a fixed parameter
value GA in the following cases: a generic non-deceptive multimodal function; a  
deceptive function; a time dependent function; a travelling salesman problem 
of $33$ cities and finally a time dependent travelling salesman problem. 
Throughout we used roulette wheel selection except at one point, which will
be explicitly mentioned, where we used tournament selection of size $5$.

The simple multimodal function is seen in Figure 2. Rather than worry about 
about how to design a real valued, continuous, double peaked function 
and then approximate it by binary numbers we simply assigned a fitness value to every
integer between $0$ and $63$. This function is not deceptive in that crossover
between optimal or near optimal strings does not produce very unfit strings, i.e.
crossover of strings near $10$ and near $40$ does not tend to produce strings between
$20$ and $30$, or between $50$ and $63$. The GA we then used was associated
with a fixed population of $500$ individuals. The basic chromosome was a 
$6$-bit binary string representing the integers between $0$ and $63$. 
In coding the mutation and/or crossover probabilities
we ranged between $3$ and $16$-bit additions to the basic chromosome depending
on whether one or both parameters were coded and whether or not they were
coded using $3$-bit or $8$-bit binary representations, the latter obviously giving
a finer representation. The performances of  various GAs with and 
without parameter codification were then tested. In comparing with the performance
of a standard GA we chose mutation and crossover probabilities of $0.01$ and $0.8$
respectively, these being the generally recommended values in the literature \cite{param}.

The second landscape we chose to investigate is shown in Figure 3. 
The landscape is deceptive in the sense that crossover between fit strings 
associated with the optima $000000=0$ and $111111=63$ produces very unfit strings.
The same tests were carried out as with the previous landscape.

The third landscape chosen included time dependence. The initial landscape is that
shown in Figure 2, which has as global optimum $10$ and $11$. However, after 60\% of the 
population reach the global optimum the landscape is suddenly changed to that of Figure 4
wherein the original global optimum is now only a local optimum and a new, very narrow
global optimum appears at $62$. We call this a ``jumper'' landscape. In tests
with the jumper landscape we used tournament selection of size 5 and also imposed
a lower bound of $0.005$ for mutation.

The above landscapes were all chosen to illustrate clearly various advantages of
parameter codification in a context where the optima are all explicitly known. 
As an example of a system where the optima are not a priori known and where the
size of the state space is very large we considered a $33$-city travelling salesman
problem (TSP). The TSP is a prototypical NP-complete problem which is easy to
state but difficult to solve. Here our aim was to compare the performance of a
GA with fixed probabilities for the genetic operators with the same GA but with 
coded probabilities, not to find the best codification and set of genetic operators 
for applying GAs to the TSP. It is known that GAs are quite competitive with other
optimization techniques \cite{homaifar} when coded appropriately.

With that in mind we chose the simplest minded codification via a path 
representation wherein the cities are listed in the order they are visited. 
The genetic operators used were: ``mutation" (permutation of two randomly selected 
cities ) and ``crossover" (inversion of the cities between two randomly selected 
points on the chromosome). For example: for a $6$ city problem mutation at $0$ 
and $3$ of the possible route $134520$ leads to ${\bf 5}34{\bf 1}20$. Inversion 
between points $2$ and $5$ of the same route leads to $13{\bf 254}0$.
We did not bother to avoid cyclic permutations of a given route as once 
again our principal aim is to compare the performance of a given GA with and 
without parameter codification. The operator probabilities were coded using
$8$ bits for each one resulting in a $49$-bit chromosome. Although two-city
interchange and inversion were used on the $33$ bits that code for the route
standard mutation and crossover were used on the remaining $16$ bits.
For mutations we set a lower bound of $0.001$ rather than zero. 

The fitness function for the $n$-city TSP, as is well known, 
is just the total distance of a route
\be
R=\sum_{i=0}^{n-1}(d(c_i,c_{i+1}))+d(c_0,c_{n-1})
\ee
with $d(x,y)$ the distance between cities $x$ and $y$, subject 
to the restrictions that all cities must be visited and no city can
be visited more than once. 

Finally, we considered the above problem but in a time dependent context
where starting with a TSP of $23$ cities at generation  $3000$ ten more cities 
were added so that a new optimum route completely different to the first
had to be found. To compare the different performances of the GAs in the
TSP we used the following function
\be
{\cal F}=\sum_{\s i=j}^{\s j+100}
{R_{\s min}(i)+R_{\s av}(i)\o R_{\s min}(i)-R_{\s av}(i)}\label{fit}
\ee
where the sum is over an interval of $100$ generations and $j$ represents
the discrete time at which we are evaluating $\cal F$. Thus over a period of
$6000$ generations we will have a set of $60$ points to evaluate $\cal F$.
$R_{\s min}(i)$ is the shortest route length found at generation $i$ and 
$R_{\s av}(i)$ is the average route length for generation $i$.  
Note that this function emphasizes more the notion of population
fitness than the best individual fitness. In terms of pure combinatoric
optimization knowing that there exists a unique optimum state it is
of course finding the fittest individual in the shortest amount of time that
is of interest. However, in many other problems, for instance in evolution
and in other problems with a time dependent environment what is 
important is fitness of the population. The above fitness function takes
into account both convergence speed and average population fitness.

\section{Results}

In Figure 5 we see the results of various GAs in the landscape of Figure 2.
The plot shows in the upper half relative frequency of the optimum string as
a function of time and in the lower half the average values of the codified 
GA parameters as a function of time. The initial population was chosen at
random. Note that the population size here is large compared to the size of
the state space as our intention was to investigate the effects of parameter codification
without strong finite size effects complicating the picture due to sampling errors. 
We will consider the latter thoroughly in a future publication. 

The most notable feature of Figure 5 is the behaviour of the coded 
parameters. For mutations we see that the system begins to ``cool'' itself down
as the population begins to order. It is clear that there is no direct selective benefit
in a given generation for one mutation rate versus another, however, we do see 
quite clearly the effects of symmetry breaking in that strings with low mutation
rates are likely to have fitter offspring. This symmetry breaking becomes more
pronounced as a function of time. With respect to crossover we see that there
is rather a tendency to increase than decrease as a function of time. One can
understand this from the point of view that crossover is strictly neutral for the
fully ordered component of the population, i.e. all the optimum strings, whilst
a high crossover rate helps in getting rid of unfit strings as found in 
\cite{stewael}. In terms of comparison with a fixed parameter GA we see that 
an $8$-bit encoded GA spends more time trying to find the 
optimum mutation and crossover rates as it has to search through more possibilities. 
Note that the steady state population for a codified mutation rate will always be superior to
that of a fixed rate simply because the population can only be strictly ordered
when $p=0$.

In Figure 6 we show what happens with the landscape in Figure 2 but now in the case 
where the initial population is totally concentrated at one point, $49$. One might
ask why we would want to consider such a case. The reason why will become 
more obvious after seeing the results from the jumper landscape. Suffice it to 
say here that in a time dependent landscape it might well be that the population 
has pretty much converged to an optimum but that at a certain point the 
landscape changes such that the original global optimum is now only a 
local optimum and the system must then seek the new global optimum. If the 
landscape changes only after the population has converged to the original 
optimum then in terms of evolution in the new landscape the system is starting from
a very special initial condition. 

Considering the explicit results we see that a 
fixed parameter GA performs particularly badly. The reason why is simple: crossover
does not act very efficiently in encouraging diversification when one
starts with an ordered population. This role has to be played by mutation. If the 
mutation rate is low then the search time to find the optimum is large. It is clear that
codifying crossover and not mutation does not help. When mutation is codified
the results are clearly far superior. Once again we see how the system cools
itself down after the GA begins to find the optimum. The behaviour of the crossover
probability as a function of time is interesting in that it shows an increase even
though we have emphasized that crossover is not very efficient. The point is, that
despite its inefficiency it is still a positive operator in that it aids the search for
the new optimum. Only in the case of zero mutation rate and a totally ordered
population is it strictly neutral.

The next figure, Figure 7, shows the results associated with the deceptive landscape 
of Figure 3. The initial population was biased with $100$ individuals being placed 
at each optimum to make the problem even more ``deceptive''. One sees that the
fixed parameter GA was incapable of maintaining the population at the global optimum.
The $8$-bit coded GA by contrast increased the relative concentration of the 
optimum without any problems. The $3$-bit coded GA sometimes converged to the
global optimum and sometimes to the local one. Once again we see that in terms
of mutations the system cools itself down. The most interesting result here however
is what happens to the crossover rate. In the results from the non-deceptive 
landscape the crossover probability rose monotonically. Here, however, we see
that initially there is a very sharp drop in the rate. This is a direct response to the
deception. Crossover of $0$ strings with $63$ strings produces very unfit results. 
As the system starts with a large number of them those that are coded with low
crossover rates will be preferred. Eventually as the system begins to order around
the global optimum crossover loses its destructive nature and so the net rate
increases.
 
Figure 8 contains the results for the deceptive landscape when $p=0$. In this case we
started with an initial population of $80$ individuals, $20$ of which were located 
at one optimum, $20$ at the other and the remaining $40$ distributed 
randomly between them. Once again, initially, due to the large populations associated with the  
two optima, crossover is very destructive as any crossover which includes strings
from both optima will result in very unfit offspring. Note that the average population 
fitness first decreases somewhat although this decrease is minimal compared to
the fitness decrease in the fixed parameter GA.

In Figure 9 we see what happens for the jumper landscape. The upper curves show
the relative frequencies of the optima using $8$-bit and $3$-bit codification and also 
what happens when $p_c=0$ and only the mutation rate is coded. There are several 
notable features: first of all we see that the fixed parameter GA was incapable of finding
the new optimum whereas the coded GAs had no problem whatsoever.  
For the case $p_c=0$ the curve $40,41$ shows the relative frequency
of the strings associated with the optimum at $40$ and $41$. Before the landscape
``jump'' this optimum is local being less fit than the global optimum 
at $10$ and $11$. After the ``jump'' it is fitter but less fit than the new 
global optimum $63$ which is an isolated point. One thus sees that
the optimum was found in a two-step process after the landscape change. First the
strings started finding the optima $40$, $41$ before moving onto the true global
optimum, $63$. Immediately after the jump the effective population of the new
global optimum is essentially zero. The number of strings associated with 
$40$ and $41$ first starts to grow substantially at the expense of $10$ and
$11$ strings. At its maximum the number of optimum strings is still very low,
however, very soon thereafter the algorithm manages to find the optimum string which
then increases very rapidly at the expense of the rest. The striking result
here can be seen by comparing the changes in the relative frequencies with
the changes in the average mutation rate, especially in the case $p_c=0$. 
Clearly they are highly correlated.
First, while the population is ordering itself around the original optimum,
there is an effective selection against high mutation rates as one can see
by the steady decay of the average mutation rate. After the jump one can
see a noticeable increase in the mutation probability as the system now has to 
try to find fitter strings. As the global optimum is an isolated state it is much 
easier to find fit strings associated with $41$ and $40$. The system now thinks
that this is the optimum and starts to cool down again only to find that this 
was not so, whereupon the system heats up again to aid the removal of the
population to the true global optimum. It is clear that there is a small delay
between the population changes and changes in the mutation rate. This 
is only to be expected given that there is no direct selective advantage 
in a given generation for a particular mutation rate. The selective advantage
of a more mutable genotype over a less mutable one
can only come about via a feedback mechanism.
It is precisly this feedback process that is described and measured by the 
effective fitness function. The average mutation rate also grows due to another
effect which is that the new optimum is more likely to be reached by strings
with high mutation rates which then grow strongly due to their selective advantage. 
Thus high $p$ strings will naturally dominate the early evolution of the global
optimum. After finding the optimum however it will become disadvantageous
to have a high mutation rate hence low mutation strings will begin to 
dominate.

We now turn to the case of the TSP. In Figure 10 we see a comparison
between GA performance for fixed rates, $p=0.1$, $p_c=0.2$, and coded rates.
One can see that in terms of convergence velocity the two are roughly 
comparable, however in terms of the algorithm fitness function, (\ref{fit}), the
coded algorithm performs much better due to the fact that the average member
of the population is almost as good as the best. The relative performances are
shown in Figure 11. The shortest route for this particular problem reported in 
\cite{tspresult} is 10930. The best result from our GA 
is $11,040$ for fixed parameters and $11,662$ for coded parameters.
We found that inversion affects the GA's performance more than mutation 
due to its more global scope. 

In Figure 12 we see the results for the jumper TSP landscape and in 
Figure 13 the efficiencies of the two GAs are compared.
Finally in Figure 14 we see the evolution of the coded crossover rate 
in the non-jumper and jumper
landscapes. Note that very soon after the landscape change at $t=3000$ the crossover rate
begins to rise reaching a peak at $t=3500$ where it is almost double its prejump value.
Thus as in the case of the previous jumper landscape one sees that 
the parameters respond to changes in the landscape. In this case it is the inversion
rate which rises after the landscape change in order to aid the algorithm in the 
search for the new optimum.

\section{Discussion and Conclusions}

In this paper we have investigated the effects of coding the probabilities
that govern the action of genetic operators in a simple GA based on
a select set of model fitness landscapes. One of the basic premises
of the paper was to investigate how parameter codification may help 
in adaptive search problems and to illuminate some of the consequent
theoretical issues. 

Optimizing the exogeneous parameters of a GA is a very tricky business
that at the end of the day must be landscape dependent. Generically
this will also mean time dependent. We have therefore tried to show what
one might gain via a codification of these parameters within the chromosomes
themselves thus leading to an essentially, totally autonomous system.
We saw that in a time-dependent landscape certain parameter values
will be favoured over others in order to facilitate the search for new optima, 
and that the values themselves will change over time. However, the chief lessons
we wish to point out from our work here have much wider implications than
those to be drawn from a simple comparison of coded versus non-coded GA
performance. 

What are these lessons? There are two: the lesson of induced 
symmetry breaking, and the lesson of effective fitness. In the former,
if we have a degenerate genotype-phenotype mapping, i.e. there exist
synonymous genotypes, then there exists a symmetry, or perhaps better
to say an equivalence relation on $G$. By degenerate
we mean explicitly that the different genotypes all correspond to exactly
the same phenotypic fitness value. The fitness landscape as a function of
genotype thus has neutral directions. In the presence of pure selection the 
system will be unable to distinguish between these directions. However, 
on including mutation and/or crossover this degeneracy will be lifted, certain
directions now being preferred over others even though there is no direct
selective advantage for them. Thus we may say there is an induced 
symmetry breaking. The toy model of section 3 and the later numerical
results all confirm this clearly and explicitly. In understanding this 
phenomenon the normal concept of fitness is of little use, which brings us to the
second important lesson, that of effective fitness. Even though 
certain directions in $G$ may be neutral in terms of fitness they most definitely
are not in terms of the effective fitness. We therefore claim that effective
fitness, and the consequent fitness landscape in terms of it, are much more 
relevant and meaningful concepts in the presence of other genetic operators
such as mutation and crossover. The effective landscape will be time dependent
even though the original landscape wasn't. We will give a more thorough analysis
of the effective fitness in a future publication.

Also of interest is the relation to the theory of natural evolving systems. There the
bare mutation rates have a biophysical origin which cannot be coded directly, however
it is also true that different synonymous codons can have varying mutabilities. The 
most obvious reason for this is that point mutation itself does not respect the 
equivalence between synonyms. For example, in the case of the six leucine codons,
{\it CTPu} have four possible silent point mutations, {\it CTPy} have three and 
{\it TTPu} have two. Therefore, if missense mutations are selected negatively 
then one will find that {\it CTPu} codons have an effective selective advantage due
to their higher resistance to mutation. Vice-versa, in the recognition regions of 
the enveloping protein of a lentivirus there is a preference for 
codons that mutate non-synonymously, in order to escape detection by
the immune system \cite{codon}. One can view this effect as an 
example of self-adaptation where the choice of mutation rate is realized indirectly
through the codon bias.

\vskip 0.1in

\subsection*{Acknowledgments}

This work was partially support by a CONACYT scholarship to J. Mora-Vargas
and by DGAPA grant IN105197.

\vfill\eject

\subsection*{Figure Captions}

\noindent Figure 1: Graph of relative frequency as a function of time. The squares 
represent genotype C, the diamonds genotype A and the crosses genotype B.
\vskip 10pt 
\noindent Figure 2: Graph of fitness versus genotype. 
\vskip 10pt 
\noindent Figure 3: Graph of fitness versus genotype for ``deceptive'' landscape.
\vskip 10pt

\noindent Figure 4: Graph of fitness versus genotype after ``jump'' from Figure 2.
\vskip 10pt 

\noindent Figure 5: Graph of relative concentration of the optimum (CR) and 
average crossover and mutation probabilities as a function of time. CR is shown
in the upper half and the average rates in the lower half. CR-FR is the curve for 
a fixed rate GA; CR-3b for coded $3$-bit probabilities; CR-8b for coded 
$3$-bit probabilities; Mut 3b, Mut 8b, Cross 3b and Cross 8b are the average
mutation and crossover probabilities in $3$-bit and $8$-bit representations.
\vskip 10pt 

\noindent Figure 6: The same as in Figure 5, except now the initial population
is centered at $49$ in the landscape of Figure 2. CR-Cro 3b represents a GA
where $p$ is fixed at $0.01$ and crossover is $3$-bit coded. Cros 3b is the 
corresponding average crossover rate.
\vskip 10pt 

\noindent Figure 7: Same curves as in Figure 5 but now for the ``deceptive'' 
landscape of Figure 3 and with different initial population.
\vskip 10pt 

\noindent Figure 8: Graph of relative concentration and 
average crossover probabilities as a function of time for the landscape 
of Figure 3 with initial population size of $80$ and $p=0$. CR-Fit.1 is the relative 
concentration of the global optimum at point $0$; CR-Fit.2 
is the relative concentration of the local optimum at point $63$;
Avg. Cross is the average crossover rate and Avg. Fit the average 
population fitness.
\vskip 10pt 

\noindent Figure 9: Graph of relative concentration of the optimum (CR) and 
average crossover and mutation probabilities as a function of time for the
``jumper'' landscape. CR-3b and CR-8b are the results for $3$-bit and $8$-bit
encoded algorithms CR-Mut8b is the result for coded mutation with $p_c=0$, 
with $40,41$ being the relative concentration of strings associated with the local
optimum at $40$ and $41$. Mut 3b, Mut 8b, Cross 3b and Cross 8b are the average
mutation and crossover probabilities in $3$-bit and $8$-bit representations. The
solid line for Mut8b is the average mutation rate in the case $p_c=0$.
\vskip 10pt 

\noindent Figure 10: Graph of Fitness (Route length) versus time for $33$-city TSP.
f2.fit and f2.bes represent the average and best fitnesses for a fixed parameter GA
whilst s2.fit and s2.bes are for a coded $8$-bit GA.
\vskip 10pt 

\noindent Figure 11: Graph of Efficiency versus time for fixed parameter and coded
parameter GAs in the TSP; f2.eff for the fixed and s2.eff for the coded GA.
\vskip 10pt 

\noindent  Figure 12: Graph of Fitness (Route length) versus time for ``jumper'' TSP.
f2.fit and f2.bes represent the average and best fitnesses for a fixed parameter GA
whilst s2.fit and s2.bes are for a coded $8$-bit GA.
\vskip 10pt 

\noindent Figure 13: Graph of Efficiency versus time for fixed parameter and coded
parameter GAs in the ``jumper'' TSP; fj2.eff for the fixed and sj2.eff for the coded GA.
\vskip 10pt 

\noindent Figure 14: Graph of Crossover rate versus time for the $33$-city TSP 
and the ``jumper'' TSP. sj2.cro is for the ``jumper'' landscape

\vfill\eject

\begin{thebibliography}{99}

\bibitem{kauffman}  S. A. Kauffman, {\it The Origins of Order}, 
Oxford University Press, Oxford (1993).

\bibitem{1} T. B\"ack, {\it Evolutionary Algorithms in Theory and Practice: 
evolution strategies, evolutionary programming, genetic algorithms\/}, 
(Oxford Univ. Press 1996).

\bibitem{fogel} D.B. Fogel, {\it System Identification through Simulated Evolution: A
Machine Learning Approach to Modeling}, (Ginn Press, Needham, MA, 1991).

\bibitem{davis} {\it Handbook of Genetic Algorithms} L. Davis, (ed.)  (Van Nostrand Reinhold,
New York 1991); Holland, J.H., {\it Scientific American}, July 1992, 66-72.

\bibitem{param} K. De Jong, {\it An Analysis of the Behaviour of a Class of Genetic
Adaptive Systems\/}, Ph.D thesis, (Univ. of Michigan, 1975); J.D. Schaffer, R.A. Carauna, 
L.J. Eshelman and R. Das, {\it Proceedings of the Third Int. Conf. on Genetic Algorithms 
and their Applications}, 51-60 (Morgan Kaufman 1989); J.J. Grefenstette, {\it IEEE 
Transactions on Systems, Man and Cybernetics\/}, SMC-16(1), 122 (1986). 

\bibitem{hart} W.E. Hart and R.K. Belew, {\it Proceedings of the 
Fourth Int. Conf. on Genetic Algorithms and their Applications}, 190-195 (Morgan Kaufman 1991).

\bibitem{schwefel} H.P. Schwefel, {\it Numerical Optimization of Computer Models\/}, (J.P. Wiley 
1981).

\bibitem{back} T. B\"ack, {\it Parallel Problem Solving from Nature 2\/}, ed.s R. M\"anner and
B. Manderick, (Elsevier 1992) 85; {\it Proceeding of the First European Conference on 
Artificial Life\/}, ed.s F.J. Varela and P. Bourgine, (MIT Press 1992) 263. 

\bibitem{mori} J.D. Schaffer and A. Morishima, {\it Proceedings of the Second Int. 
Conf. on Genetic Algorithms and their Applications}, 36-40 (Lawrence Erlbaum Associates 1987)

\bibitem{stewael} C.R. Stephens and H. Waelbroeck, {\it Analysis of the
Effective Degrees of Freedom in Genetic Algorithms }, ICN, UNAM preprint:
ICN-UNAM-96-08; adap-org/9611005; {\it Proceedings of the 
Sixth Int. Conf. on Genetic Algorithms and their Applications},  (Morgan Kaufman 1997).

\bibitem{goldberg} D. E. Goldberg, {\it Genetic Algorithms in search, 
optimization and machine learning}, Addison Wesley, Reading, MA (1989).

\bibitem{Wright} A.H. Wright, {\it Foundations of Genetic Algorithms\/}, 205 
(Morgan Kaufmann 1991).

\bibitem{Caruna} R.A. Caruna and J.D. Schaffer, {\it Proceedings of the 5th Int. Conf.
on Machine Learning\/}, 153 (Morgan Kaufmann 1988).

\bibitem{Kimura} M. Kimura, {\it The Neutral Theory of Molecular
Evolution}, Cambridge University Press, Cambridge (1983).

\bibitem{homaifar} A. Homaifar, S. Guan and G.E. Liepins, {\it Proceedings of  the 
?? Int. Conf. on Genetic Algorithms and their Applications},  (Morgan Kaufman 198?)

\bibitem{tspresult} R. L. Karg and G. L. Thompson, {\it Management Science\/}\ {\bf 10}, 225 (1964). 

\bibitem{codon} H. Waelbroeck, {\it Codon Bias and Mutability in HIV Sequences}, National
University of Mexico Preprint ICN-UNAM-97-09 (adap-org/9707) (1997).
 
\end{thebibliography}
\end{document}